\documentclass[10pt,journal,letterpaper]{IEEEtran}
\usepackage[T1]{fontenc}
\usepackage{newtxtext,newtxmath}
\usepackage{graphicx,array,cite,url,balance}
\usepackage{tikz,pgfplots}
\pgfplotsset{compat=1.18}
\definecolor{ink}{HTML}{263238}
\definecolor{ocean}{HTML}{0072B2}
\definecolor{green}{HTML}{009E73}
\definecolor{amber}{HTML}{D55E00}
\definecolor{purple}{HTML}{7A5195}
\begin{document}
\title{Wireless Networks That See: Integrated Sensing and Communication in Complex Environments}
\author{Xingqin~Lin}
\markboth{}{}
\maketitle
\begin{abstract}
The same radio interactions that complicate wireless communication can reveal what is happening around a network. Integrated sensing and communication (ISAC) turns these interactions into opportunities to observe unmanned aerial vehicles (UAVs), road vehicles, and industrial mobile robots. Yet a strong echo is not necessarily a reliable object observation: stationary background can conceal motion, reflections can create virtual objects, and one viewpoint can leave part of an object's motion unobserved. This tutorial explains how known cellular signals are used to estimate propagation, suppress stationary background, interpret reflected paths, and combine observations from different sites. UAV examples under a Third Generation Partnership Project (3GPP) stochastic channel model and ray tracing in a San Francisco urban scene illustrate the mechanisms through spatial layouts, channel profiles, and before/after range--Doppler maps. Local checks reduce false outputs, with some loss of detection. Cooperation improves position accuracy for UAVs detected by both a single site and cooperating sites. The examples show how signal resources, measurement reporting, and application needs shape sensing performance, providing practical guidance for sensing services in complex environments.
\end{abstract}
\section{Wireless Networks as Environmental Sensors}
A radio transmission interacts with the world on its way to a receiver. Buildings redirect it, moving objects change its phase, and antenna arrays resolve differences in its direction. Integrated sensing and communication (ISAC) uses these interactions to carry data and observe the surroundings through shared infrastructure. Such sensing could locate an unexpected unmanned aerial vehicle (UAV) in an air corridor, monitor traffic at a road intersection, or track industrial mobile robots in factories and warehouses. These applications make environmental observation an important direction for sixth-generation (6G) networks and beyond~\cite{lin_bridge6g,liu_isac}.

The observation required by an application determines the engineering problem. Detecting whether objects have entered or left an area can use signal changes from many overlapping reflections, as illustrated by Wi-Fi sensing~\cite{wu_wifi}. Locating an object requires interpreting measured delays and directions as a location. Following a vehicle requires repeated observations and timely updates. Positioning a device that transmits a known signal differs from sensing an object that need not transmit its own signal. A channel estimate that supports reliable symbol detection does not, by itself, identify which object produced each path.

ISAC offers several architectures. A radio site can illuminate and receive its own echoes, separated sites can share the transmitting and receiving roles, and mobile platforms can bring sensing viewpoints to new locations~\cite{meng_uav}. Each architecture exposes different path lengths, interference conditions, and synchronization requirements. Their common foundation is a sequence of decisions: obtain informative radio measurements, separate desired returns from background, interpret candidate paths, and combine observations when one view is insufficient.

This tutorial follows that sequence through UAV examples in a Third Generation Partnership Project (3GPP) stochastic environment and a San Francisco ray-tracing environment. Layouts, channel profiles, and sensing maps connect propagation and receiver processing to broader lessons for sensing design and measurement reporting.

\section{Reading the Environment from Reference Signals}
\subsection{Three Dimensions of a Radio Observation}
Reference signals place known symbols at specified frequencies and times. Dividing the received reference symbol by its known transmitted value estimates how propagation has changed its amplitude and phase. Repeated estimates across frequency, time, and antenna elements expose three complementary aspects of propagation~\cite{sturm_ofdm}. Variation across frequency reveals travel delay. Phase evolution over time reveals the rate of change of path length, observed as a Doppler frequency shift. Differences across antenna elements reveal arrival direction. Together, they constrain an object's position and motion.

The physical interpretation depends on the path. In monostatic sensing, transmission and reception occur at the same site. A direct target return travels to the object and back; half its traveled distance gives range. Doppler measures motion toward or away from that site, called radial speed. An object moving sideways can have little radial speed even when its total speed is large. In bistatic sensing, the transmitter and receiver are at different locations. The measured path length is the sum of their distances to the object. Their positions and timing offsets must be known to interpret that sum. A reflected path measures its complete route, and its arrival direction may point toward a wall rather than toward the object.

Bandwidth separates nearby delays; observation duration separates nearby Doppler shifts; antenna aperture, the array's physical extent, separates nearby directions. These resources serve different purposes. More bandwidth cannot resolve a directional ambiguity, and changing motion can prevent echoes from reinforcing one another over a longer observation. Polarization, the orientation of the radio wave's electric field, provides another view of scattering. Array weights determine which directions are illuminated and observed. A narrow beam can strengthen one return while poorly illuminating another object.

\subsection{A Cellular Reference-Signal Example}
Positioning reference signals (PRS) in fifth-generation (5G) New Radio (NR) illustrate how known cellular symbols can probe an environment~\cite{wei_prs}. The case-study assumptions largely follow the UAV-sensing evaluation framework of the 3GPP Release 20 study on ISAC for NR~\cite{tr38765}. The case study uses a 4 GHz carrier, 30 kHz spacing between frequency tones (subcarriers), and 273 carrier resource blocks, each containing 12 tones. A full-band sensing site occupies 272 blocks. A slot lasts 0.5 ms and contains 14 symbols. Each PRS transmission occasion contains three resources, each a four-symbol pilot group on half the tones.

Each resource sends one known sequence on a logical antenna port; antenna weights distribute it across physical elements. The three resources use different known beamforming weights, which set each element's amplitude and phase. Their relative responses form a transmit-direction signature, complementing receive-array information. An eight-by-eight array receives two polarizations at each element, providing 128 receive ports. Across the three resources, these give 384 resource--port measurement combinations, called coded receive channels. The simulation samples the channel at reference-symbol frequencies and times, assuming ideal timing and separation between tones.

Thirty-two occasions, spaced by 4 ms, form a nominal 128 ms coherent processing interval (CPI). Coherent processing first adjusts each echo's phase for a candidate motion, then adds the echoes with their phases retained. A consistent return reinforces itself. A longer observation helps distinguish nearby Doppler shifts, but gaps between occasions allow different speeds to produce the same sampled phase change. At 4 GHz and 4 ms spacing, radial speeds separated by about 9.37 m/s are indistinguishable from these samples alone. Closely spaced symbol pairs within each resource provide a second time scale to resolve this speed ambiguity.

The arrangement occupies approximately 5.34\% of the carrier's time--frequency resource elements: 272 of 273 blocks, 12 of 14 symbols, half the tones, and one slot in every eight. A resource element is one tone during one symbol. Sensing also requires processing, reporting, and coordination. Existing communication or positioning pilots can be reused if their arrangement supports the observation. Extra transmissions compete with data for airtime and energy; beams directed toward a target may serve a communication user poorly.

\begin{figure}[t]
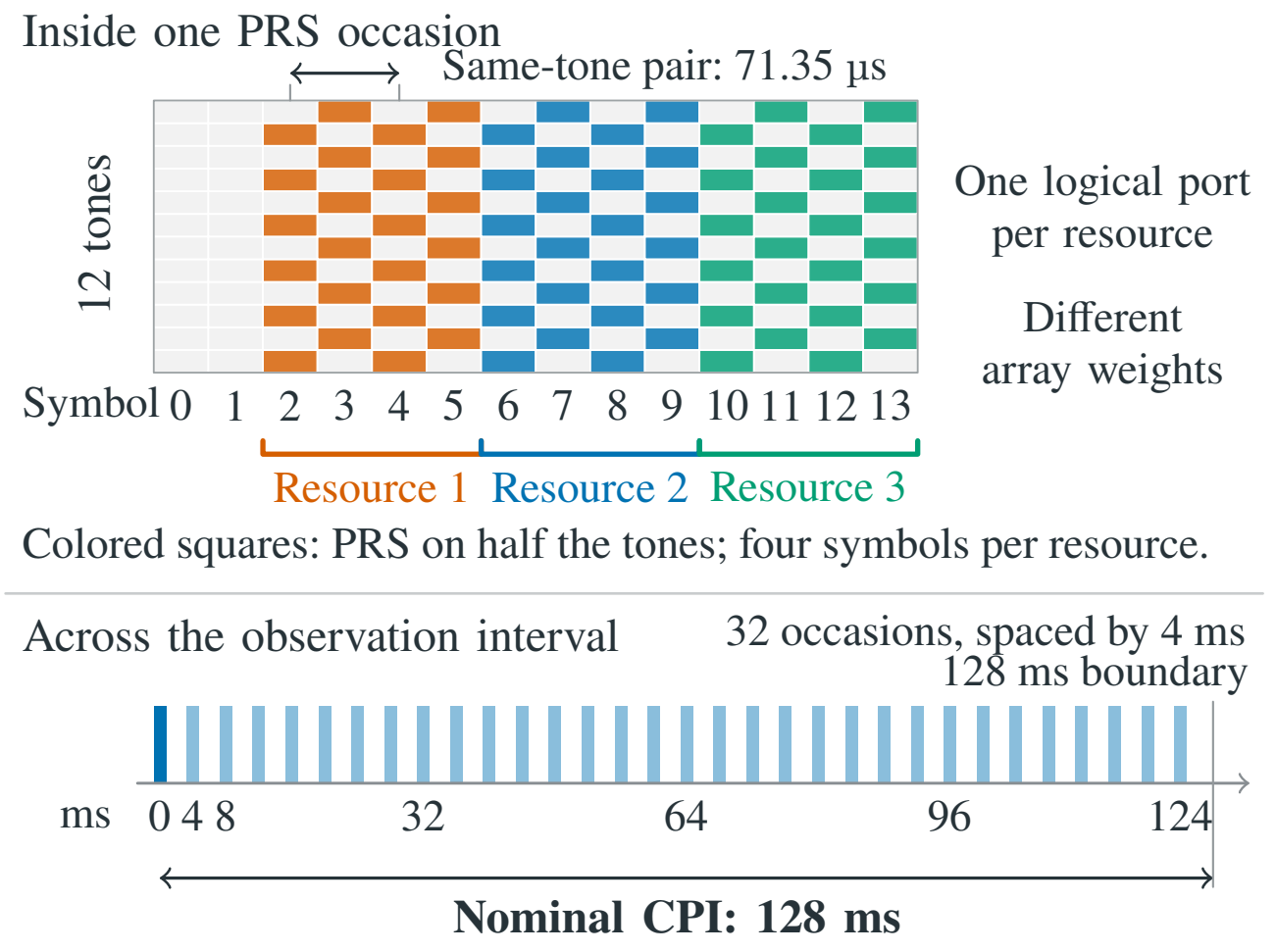

\centering



\caption{\textbf{PRS timing.} Three one-port resources use different array weights. Colored cells show pilots on alternating tones (comb-2); short symbol pairs help resolve speed ambiguity. Occasion starts span 0--124 ms within the nominal 128 ms CPI.}
\label{fig:waveform}
\end{figure}

The occupied 97.92 MHz bandwidth gives an ideal monostatic range resolution of about 1.5 m. The 128 ms CPI places neighboring Doppler bins about 0.29 m/s apart. Windowing, which weights samples before processing, and echo strength affect actual peak separation. More points on a plotting grid do not improve physical resolution. For example, a radial speed of 10 m/s can appear near 0.63 m/s when sampled every 4 ms; short symbol pairs help resolve the ambiguity. Figure~\ref{fig:waveform} connects these time scales to the reference-signal arrangement. Shortening the CPI improves response time but makes nearby Doppler shifts harder to distinguish~\cite{sturm_ofdm}. UAV altitudes range from 25 to 300 m, with horizontal speeds from 0 to 50 m/s. Positions remain fixed within each CPI while velocity determines Doppler phase; acceleration and within-CPI range changes are not modeled. Inter-site interference and moving clutter are excluded.

\begin{figure*}[t]
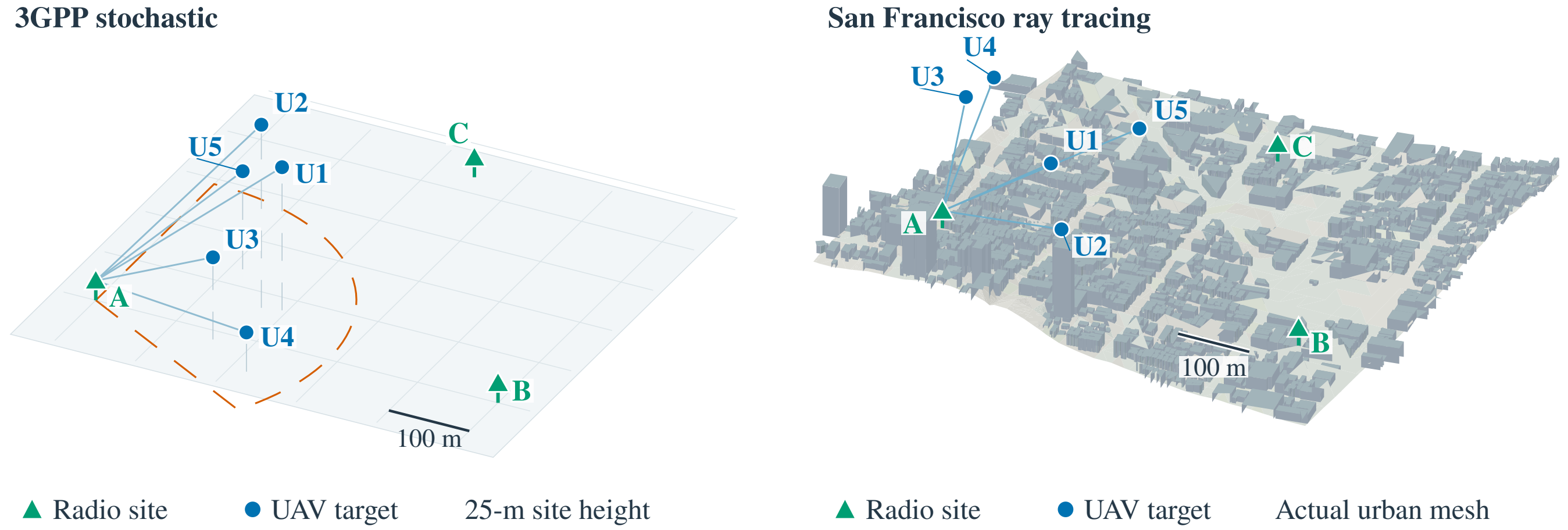

\centering
\begin{minipage}[t]{.49\textwidth}
\centering

%
\makeatother

\end{minipage}
\caption{\textbf{Propagation environments.} Independent stochastic and San Francisco drops, each with three sites and five UAVs. Lines indicate geometric sightlines; scale bars represent 100 m. Mapped geometry: OpenStreetMap contributors, through~\cite{sf_scene}.}
\label{fig:environments}
\end{figure*}

\section{Two Complementary Views of Propagation}
\subsection{Target Returns, Background, and Self-Interference}
A sensing channel contains paths involving the object of interest and paths involving the surroundings. Target returns include direct and reflected routes to the object. Buildings and stationary surfaces create background echoes, called clutter, that can be much stronger than the return from a small moving target. Radar cross section (RCS) expresses the strength of an object's radio scattering as an equivalent area. It is not simply the object's physical size: frequency, viewing angle, shape, material, and polarization affect how much incident energy returns toward a receiver. A weak echo can consequently result from unfavorable scattering as well as distance or obstruction.

Monostatic operation introduces another component: leakage from the local transmitter into its receiver. The receiver must preserve a weak echo beside a much stronger local transmission. Leakage that overloads the radio front end or analog-to-digital converter destroys information that digital subtraction cannot reconstruct. Isolation between transmitter and receiver and analog cancellation reduce leakage before digitization. Digital cancellation subtracts predictable leakage. Constant leakage and random residual interference need different treatment. Subtracting a constant channel component suppresses stationary returns but cannot remove random interference samples. Oscillator phase fluctuations and calibration errors can leave time-varying residuals. The case study models residual self-interference as Gaussian noise with power normalized to -94 dBm per 100 MHz, alongside thermal noise with a 5 dB receiver noise figure. This model applies after overload prevention.

\subsection{A Statistical Environment}
The stochastic view in Fig.~\ref{fig:environments} illustrates the statistical evaluation geometry. Three transmission/reception points (TRPs) provide separated views of five UAVs. The channel generates urban macrocell links to aerial vehicles using 3GPP assumptions~\cite{tr36777}. Each TRP--UAV pair is randomly assigned a line-of-sight (LOS) or obstructed condition, with probabilities determined by distance and altitude. The outgoing and returning monostatic paths share that condition. Groups of paths, called clusters, have statistical delays, directions, attenuation, and fading without specifying individual reflecting buildings.

The target channel pairs modeled routes to the object with routes back to the site, following 3GPP Technical Report (TR) 38.901~\cite{tr38901}. Adding the outgoing and return delays gives each echo's total delay. Scattering and antenna responses determine their combined complex amplitude. Three random virtual reference points generate independent obstructed urban macrocell links. Assigned delay shifts are applied to their responses, which are added as complex signals to form stationary background. These points do not represent three actual reflecting objects. TR 38.765 motivates the UAV evaluation and its performance measures~\cite{tr38765}; statistical ISAC channel research also illustrates the value of cluster-based construction~\cite{zhang_channel}. Statistical models support controlled studies over independent drops, each representing one simulated scene with its targets and channels. Visibility, height, scattering, and receiver choices can vary. Sensing also depends on how paths vary across nearby locations and how target and background channels relate.

\subsection{A Mapped Urban Environment}
The mapped view in Fig.~\ref{fig:environments} shows the alternative: tracing propagation through a three-dimensional San Francisco building and terrain map (mesh) derived from OpenStreetMap geometry~\cite{sf_scene}. Buildings and terrain determine which modeled routes are visible and where reflections occur. Sites are placed 25 m above their local terrain. The same receiver processes geometrically generated target returns, including direct and reflected paths. The traced target returns include up to two urban reflections; diffraction, diffuse scattering, refraction, and moving clutter are absent. Surface responses and target RCS remain model inputs even when geometry is explicit.

The environments answer different questions. Statistical channels reveal performance across controlled populations; a mapped scene exposes persistent obstructions and reflections. Map geometry does not specify every electromagnetic response, and the evaluated populations also differ in terrain and scattering assumptions. Neither model is measured ground truth.

\subsection{What the Channel Profiles Reveal}
\begin{figure}[t]
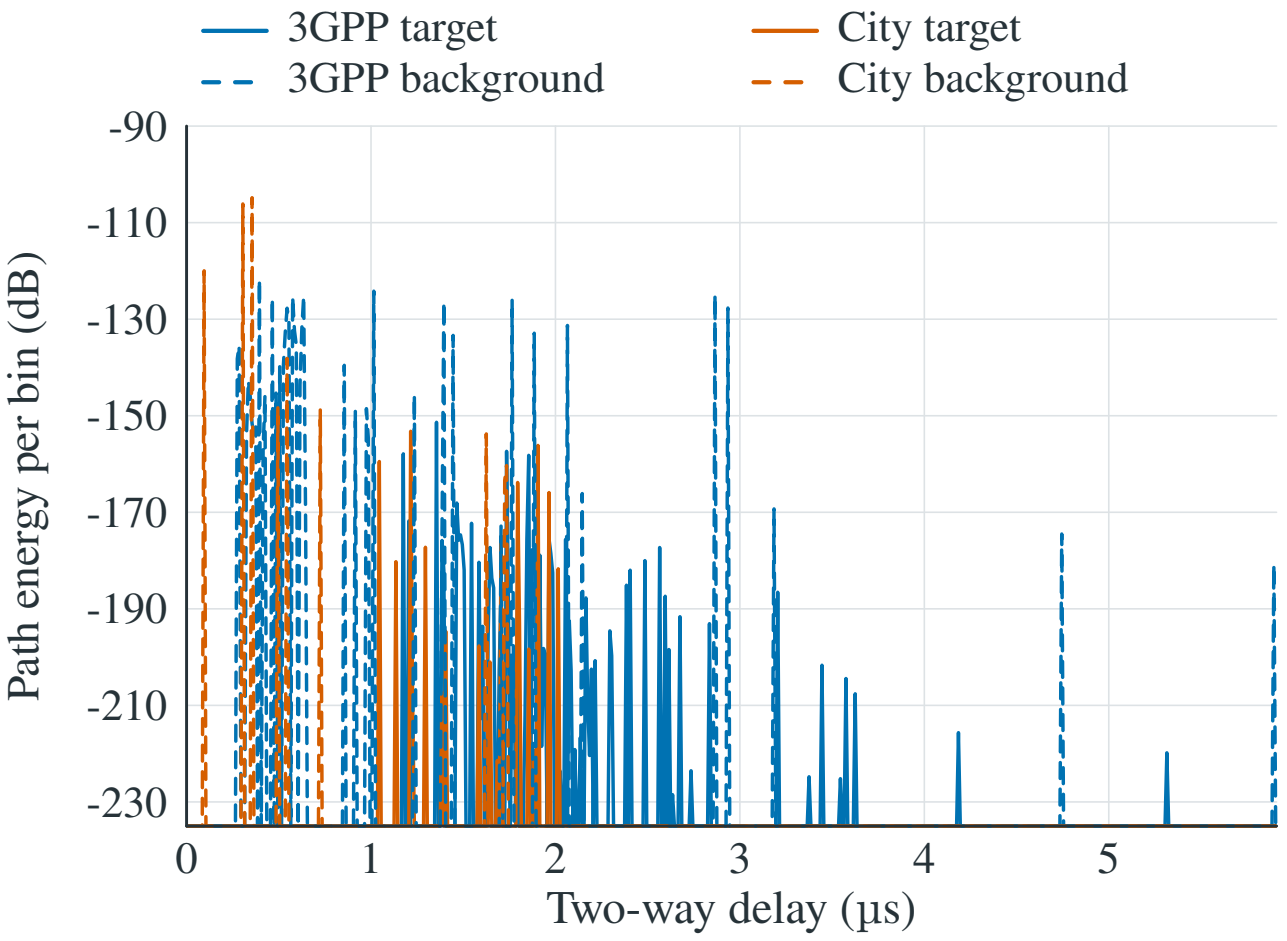

\centering
\definecolor{mrc0}{RGB}{221,226,229}
\definecolor{mrc1}{RGB}{38,50,56}
\definecolor{mrc2}{RGB}{0,114,178}
\definecolor{mrc3}{RGB}{213,94,0}

\caption{\textbf{Channel profiles.} Path energies summed in 10 ns bins and averaged over 384 coded receive channels; transmitted power is excluded.}
\label{fig:profiles}
\end{figure}

Figure~\ref{fig:profiles} compares target and background path-energy profiles in full-band snapshots. Delay gives each path's arrival time. The squared magnitude of its complex channel amplitude gives its energy contribution before different paths are combined. Statistical clusters spread energy across component paths, while mapped geometry ties routes to particular surfaces. Target and background delays can overlap, so delay alone cannot isolate an object. The profiles sum path energies in delay bins and average over the same number of coded receive channels; they do not show received power in dBm. Waves arriving at the same delay reinforce or cancel according to their phases. Thus these profiles can differ from measured spectra, and one snapshot cannot establish a general power ranking.

\section{From Echoes to Range--Doppler Peaks}
\subsection{Removing the Stationary Component}
The receiver estimates the channel at each reference symbol and subtracts its complex average across occasions. The subtraction is performed separately for each transmit resource, receive port, tone, and earlier or later measurement in a short symbol pair. Components with unchanged amplitude and phase through the CPI are removed. Subtracting complex samples preserves the phase needed for cancellation; subtracting average power would not perform the same operation. Processing across frequency yields delay, and processing across occasions yields Doppler. The resulting range--Doppler map displays return strength against apparent monostatic range and radial speed. Hann weighting gives less weight to samples near the observation edges. It reduces sidelobes, smaller peaks around a strong return that can hide weaker echoes, but broadens the main peaks. The receiver searches for candidates above a noise-dependent threshold and estimates their directions from the array.

Searching more combinations of delay and Doppler gives noise more chances to produce a false peak. The receiver sets its threshold to control this risk over the entire search region, under the assumed noise model. A coherent reflection can still cross the threshold without corresponding to a separate object.

\begin{figure}[t]
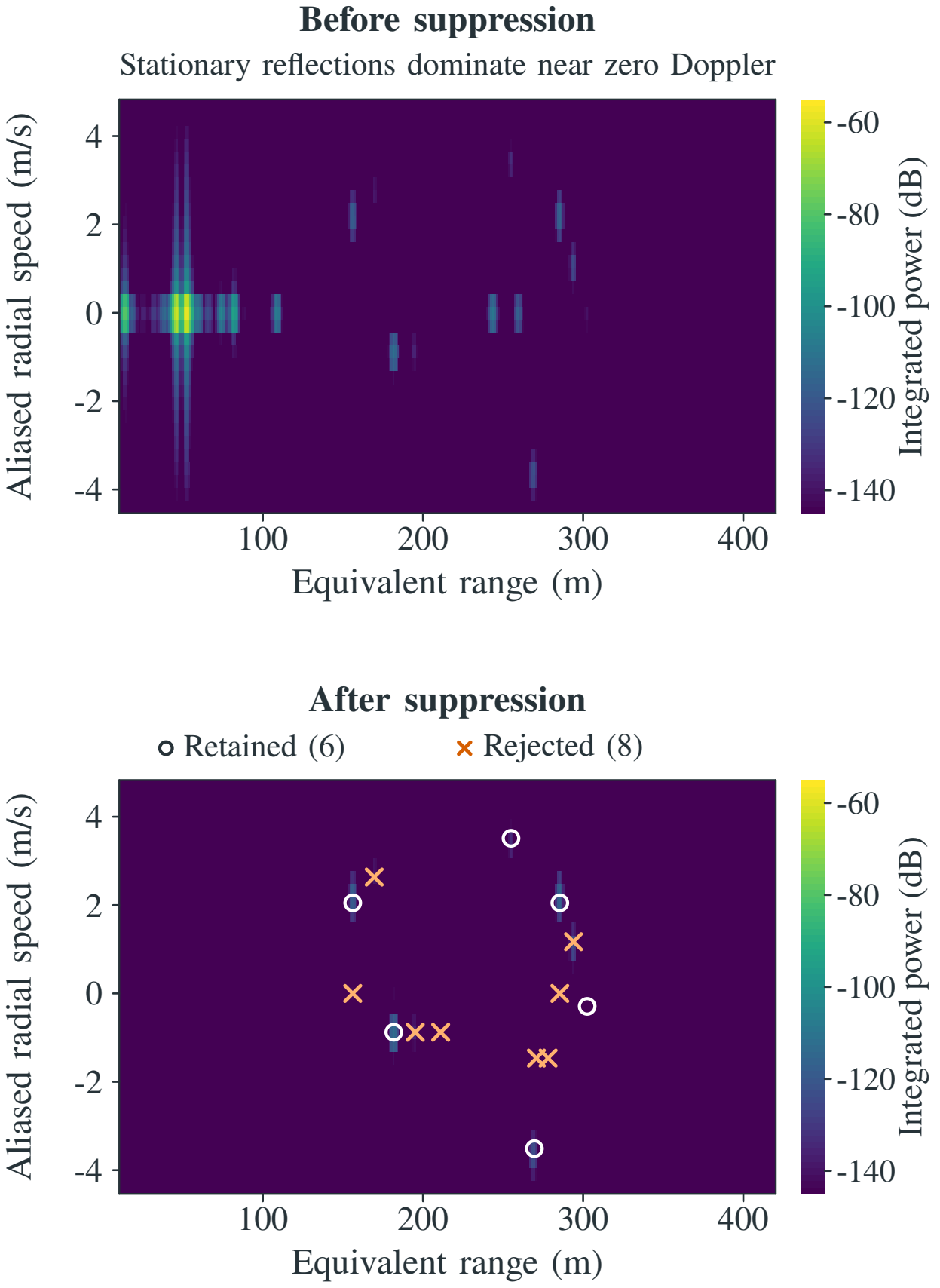

\centering
\definecolor{mrc0}{RGB}{68,1,84}
\definecolor{mrc1}{RGB}{69,7,89}
\definecolor{mrc2}{RGB}{70,18,100}
\definecolor{mrc3}{RGB}{69,12,94}
\definecolor{mrc4}{RGB}{71,29,110}
\definecolor{mrc5}{RGB}{72,34,115}
\definecolor{mrc6}{RGB}{71,23,105}
\definecolor{mrc7}{RGB}{70,45,122}
\definecolor{mrc8}{RGB}{62,74,136}
\definecolor{mrc9}{RGB}{69,50,125}
\definecolor{mrc10}{RGB}{60,78,137}
\definecolor{mrc11}{RGB}{68,55,128}
\definecolor{mrc12}{RGB}{71,40,119}
\definecolor{mrc13}{RGB}{67,60,130}
\definecolor{mrc14}{RGB}{64,70,135}
\definecolor{mrc15}{RGB}{66,65,133}
\definecolor{mrc16}{RGB}{57,87,139}
\definecolor{mrc17}{RGB}{59,83,138}
\definecolor{mrc18}{RGB}{53,95,141}
\definecolor{mrc19}{RGB}{55,91,140}
\definecolor{mrc20}{RGB}{49,103,141}
\definecolor{mrc21}{RGB}{51,99,141}
\definecolor{mrc22}{RGB}{44,115,142}
\definecolor{mrc23}{RGB}{46,111,142}
\definecolor{mrc24}{RGB}{42,119,142}
\definecolor{mrc25}{RGB}{48,107,141}
\definecolor{mrc26}{RGB}{38,131,141}
\definecolor{mrc27}{RGB}{39,127,141}
\definecolor{mrc28}{RGB}{37,135,141}
\definecolor{mrc29}{RGB}{34,143,140}
\definecolor{mrc30}{RGB}{33,154,137}
\definecolor{mrc31}{RGB}{33,150,138}
\definecolor{mrc32}{RGB}{34,161,134}
\definecolor{mrc33}{RGB}{57,183,119}
\definecolor{mrc34}{RGB}{34,165,133}
\definecolor{mrc35}{RGB}{33,147,139}
\definecolor{mrc36}{RGB}{62,187,115}
\definecolor{mrc37}{RGB}{139,213,70}
\definecolor{mrc38}{RGB}{171,219,50}
\definecolor{mrc39}{RGB}{128,210,77}
\definecolor{mrc40}{RGB}{51,180,122}
\definecolor{mrc41}{RGB}{160,217,56}
\definecolor{mrc42}{RGB}{192,223,38}
\definecolor{mrc43}{RGB}{182,221,43}
\definecolor{mrc44}{RGB}{93,199,98}
\definecolor{mrc45}{RGB}{34,158,136}
\definecolor{mrc46}{RGB}{35,139,140}
\definecolor{mrc47}{RGB}{46,176,125}
\definecolor{mrc48}{RGB}{84,196,103}
\definecolor{mrc49}{RGB}{202,225,38}
\definecolor{mrc50}{RGB}{76,194,108}
\definecolor{mrc51}{RGB}{233,228,37}
\definecolor{mrc52}{RGB}{223,227,37}
\definecolor{mrc53}{RGB}{40,172,128}
\definecolor{mrc54}{RGB}{41,123,142}
\definecolor{mrc55}{RGB}{38,50,56}
\definecolor{mrc56}{RGB}{35,169,131}
\definecolor{mrc57}{RGB}{67,191,112}
\definecolor{mrc58}{RGB}{101,202,93}
\definecolor{mrc59}{RGB}{110,205,88}
\definecolor{mrc60}{RGB}{119,208,83}
\definecolor{mrc61}{RGB}{150,215,63}
\definecolor{mrc62}{RGB}{212,226,38}
\definecolor{mrc63}{RGB}{243,230,37}
\definecolor{mrc64}{RGB}{253,231,37}


\par\vspace{1.2mm}
\definecolor{mrc0}{RGB}{68,1,84}
\definecolor{mrc1}{RGB}{71,29,110}
\definecolor{mrc2}{RGB}{69,7,89}
\definecolor{mrc3}{RGB}{70,45,122}
\definecolor{mrc4}{RGB}{62,74,136}
\definecolor{mrc5}{RGB}{69,50,125}
\definecolor{mrc6}{RGB}{60,78,137}
\definecolor{mrc7}{RGB}{68,55,128}
\definecolor{mrc8}{RGB}{72,34,115}
\definecolor{mrc9}{RGB}{69,12,94}
\definecolor{mrc10}{RGB}{57,87,139}
\definecolor{mrc11}{RGB}{51,99,141}
\definecolor{mrc12}{RGB}{55,91,140}
\definecolor{mrc13}{RGB}{46,111,142}
\definecolor{mrc14}{RGB}{49,103,141}
\definecolor{mrc15}{RGB}{64,70,135}
\definecolor{mrc16}{RGB}{71,23,105}
\definecolor{mrc17}{RGB}{53,95,141}
\definecolor{mrc18}{RGB}{70,18,100}
\definecolor{mrc19}{RGB}{67,60,130}
\definecolor{mrc20}{RGB}{66,65,133}
\definecolor{mrc21}{RGB}{71,40,119}
\definecolor{mrc22}{RGB}{59,83,138}
\definecolor{mrc23}{RGB}{38,50,56}
\definecolor{mrc24}{RGB}{48,107,141}
\definecolor{mrc25}{RGB}{44,115,142}
\definecolor{mrc26}{RGB}{42,119,142}
\definecolor{mrc27}{RGB}{41,123,142}
\definecolor{mrc28}{RGB}{39,127,141}
\definecolor{mrc29}{RGB}{38,131,141}
\definecolor{mrc30}{RGB}{37,135,141}
\definecolor{mrc31}{RGB}{35,139,140}
\definecolor{mrc32}{RGB}{34,143,140}
\definecolor{mrc33}{RGB}{33,147,139}
\definecolor{mrc34}{RGB}{33,150,138}
\definecolor{mrc35}{RGB}{33,154,137}
\definecolor{mrc36}{RGB}{34,158,136}
\definecolor{mrc37}{RGB}{34,161,134}
\definecolor{mrc38}{RGB}{34,165,133}
\definecolor{mrc39}{RGB}{35,169,131}
\definecolor{mrc40}{RGB}{40,172,128}
\definecolor{mrc41}{RGB}{46,176,125}
\definecolor{mrc42}{RGB}{51,180,122}
\definecolor{mrc43}{RGB}{57,183,119}
\definecolor{mrc44}{RGB}{62,187,115}
\definecolor{mrc45}{RGB}{67,191,112}
\definecolor{mrc46}{RGB}{76,194,108}
\definecolor{mrc47}{RGB}{84,196,103}
\definecolor{mrc48}{RGB}{93,199,98}
\definecolor{mrc49}{RGB}{101,202,93}
\definecolor{mrc50}{RGB}{110,205,88}
\definecolor{mrc51}{RGB}{119,208,83}
\definecolor{mrc52}{RGB}{128,210,77}
\definecolor{mrc53}{RGB}{139,213,70}
\definecolor{mrc54}{RGB}{150,215,63}
\definecolor{mrc55}{RGB}{160,217,56}
\definecolor{mrc56}{RGB}{171,219,50}
\definecolor{mrc57}{RGB}{182,221,43}
\definecolor{mrc58}{RGB}{192,223,38}
\definecolor{mrc59}{RGB}{202,225,38}
\definecolor{mrc60}{RGB}{212,226,38}
\definecolor{mrc61}{RGB}{223,227,37}
\definecolor{mrc62}{RGB}{233,228,37}
\definecolor{mrc63}{RGB}{243,230,37}
\definecolor{mrc64}{RGB}{253,231,37}
\definecolor{mrc65}{RGB}{255,255,255}
\definecolor{mrc66}{RGB}{255,178,111}
\definecolor{mrc67}{RGB}{213,94,0}


\caption{\textbf{Clutter suppression.} The same noisy observation and scales before and after subtraction of the complex mean. Circles mark six retained candidates; crosses mark eight rejected candidates. Checks change reports, not the plotted spectrum. The speed axis retains the sampling ambiguity.}
\label{fig:clutter}
\end{figure}

Figure~\ref{fig:clutter} shows the same noisy San Francisco observation before and after stationary-component subtraction, with identical axes and strength scales. Color denotes integrated channel-coefficient power in dB, not received power in dBm. The strong zero-Doppler ridge in the first map comes from channel components whose amplitude and phase stay constant. The 4 ms spacing makes different speeds appear at the same value on this axis; short symbol pairs resolve these alternatives later.

\subsection{The Cost of Suppression}
Any echo that appears unchanged throughout the CPI is suppressed. A stationary target is removed with stationary background, and a moving object viewed almost sideways can have near-zero radial speed and a weakened echo. A moving return can also be weakened if its phase nearly repeats every 4 ms and appears almost stationary at the sampled times. Changing foliage, vehicles, or machinery can produce moving background that survives the subtraction. Random residual self-interference adds to the background noise level. Thus some targets disappear and some clutter remains.

This selectivity connects receiver design to applications. A motion alarm may welcome suppression of time-constant responses. Monitoring a parked vehicle or a stationary obstruction requires a different strategy, such as comparing new measurements with a previously recorded, aligned background. Longer observations better separate Doppler shifts but increase delay and sensitivity to changes in motion during the measurement. Receiver processing should match the object state to be observed.

\section{From Peaks to Credible Objects}
\subsection{Consistency Across Antennas, Resources, and Time}
A spectrum peak suggests a possible propagation path. Reflection can increase its apparent range and redirect its arrival bearing. A receiver that interprets it as a direct echo can create a displaced estimate or a virtual object, a familiar multipath problem in radar~\cite{zheng_ghost}. Even a sharp, high-power peak therefore requires physical interpretation.

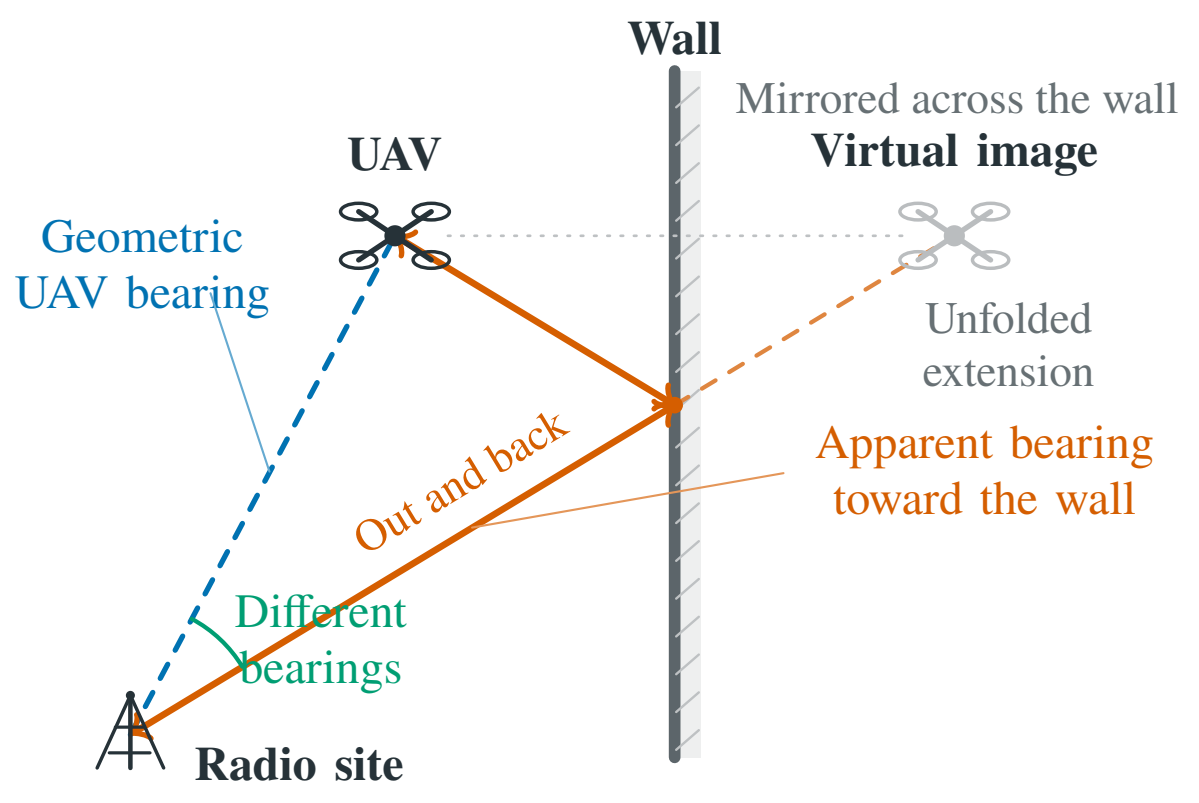
\begin{figure}[t]
\centering

\begin{tikzpicture}[x=1cm,y=1cm,line cap=round,line join=round,
  every node/.style={font=\fontsize{9}{10.5}\selectfont,text=ink}]
\path[use as bounding box] (0,0) rectangle (8.75,5.30);
\begin{scope}[yshift=-.7cm]
\fill[ink!8] (4.60,1.05) rectangle (4.77,5.59);
\draw[ink!75,line width=2.2pt] (4.60,1.05) -- (4.60,5.59);
\foreach \yy in {1.1,1.35,...,5.55} {\draw[ink!30,line width=.45pt] (4.61,\yy) -- (4.76,{\yy+.13});}
\node[font=\fontsize{9}{10.5}\selectfont\bfseries] at (4.60,5.81) {Wall};
\draw[ocean,dashed,line width=1.0pt] (1.00,1.20) -- (2.75,4.50);
\node[text=ocean,align=center] at (1.08,4.28) {Geometric\\UAV bearing};
\draw[ocean!60,line width=.4pt] (1.53,4.12) -- (1.92,2.94);
\draw[<->,amber,line width=1.25pt] (1.00,1.20) -- (4.60,3.379817);
\draw[<->,amber,line width=1.25pt] (4.60,3.379817) -- (2.75,4.50);
\draw[amber!75,dashed,line width=.8pt] (4.60,3.379817) -- (6.45,4.50);
\fill[amber] (4.60,3.379817) circle (.055);
\node[text=amber,align=center] at (6.65,2.95) {Apparent bearing\\toward the wall};
\draw[amber!65,line width=.4pt] (5.32,2.93) -- (3.26,2.57);
\node[text=amber,rotate=31.196,font=\fontsize{8.5}{10}\selectfont] at (3.18,2.88) {Out and back};
\node[font=\fontsize{8.5}{10}\selectfont,text=ink!70,align=center] at (6.81,3.78) {Unfolded\\extension};
\draw[green,line width=.8pt] (1.74,1.65) arc[start angle=31.196,end angle=62.061,radius=.865];
\node[text=green,align=center] at (2.26,1.80) {Different\\bearings};
\foreach \xx/\shade/\lab in {2.75/100/UAV,6.45/32/Virtual image} {
  \draw[ink!\shade,line width=1.0pt] ({\xx-.24},4.34) -- ({\xx+.24},4.66);
  \draw[ink!\shade,line width=1.0pt] ({\xx-.24},4.66) -- ({\xx+.24},4.34);
  \fill[ink!\shade] (\xx,4.50) circle (.07);
  \foreach \dx/\dy in {-.24/-.16,-.24/.16,.24/-.16,.24/.16} {
    \draw[ink!\shade,line width=.65pt] ({\xx+\dx},{4.50+\dy}) ellipse[x radius=.12,y radius=.055];
  }
  \node[font=\fontsize{9}{10.5}\selectfont\bfseries] at (\xx,5.04) {\lab};
}
\draw[ink!30,dotted,line width=.5pt] (3.10,4.50) -- (6.12,4.50);
\node[font=\fontsize{8.5}{9.5}\selectfont,text=ink!70] at (6.47,5.41) {Mirrored across the wall};
\draw[ink,line width=.85pt] (1.00,.98) -- (1.00,1.46);
\draw[ink,line width=.65pt] (.78,.98) -- (1.00,1.46) -- (1.22,.98);
\draw[ink,line width=.6pt] (.86,1.08) -- (1.13,1.08);
\draw[ink,line width=.6pt] (.90,1.25) -- (1.09,1.25);
\fill[ink] (1.00,1.46) circle (.03);
\node[anchor=west,font=\fontsize{9}{10.5}\selectfont\bfseries] at (1.30,1.01) {Radio site};
\end{scope}
\end{tikzpicture}

\caption{\textbf{Multipath geometry.} A conceptual wall reflection on the outgoing and return routes creates an apparent UAV at its mirror-image location. The dashed blue line points toward the actual UAV.}
\label{fig:paths}
\end{figure}

A flat wall illustrates the ambiguity in Fig.~\ref{fig:paths}. When both the outgoing and return routes reflect from the wall, their total length can equal the direct round-trip distance to the UAV's mirror image. The arrival bearing then points toward that virtual location. A stable route can remain coherent across antennas and time. Known wall geometry permits a reflected-path interpretation; a different viewpoint may supply a conflicting direct bearing~\cite{zheng_ghost}.

Spatial coherence tests whether one arrival direction explains the amplitudes and phases across receive antennas. The receiver fits the expected array pattern and compares the energy it explains with total measured energy. Amplitude and phase are fitted separately for each of the six resource/polarization combinations. Transmit-signature consistency predicts the three relative echo responses from the beamforming weights, assuming departure near the measured arrival direction. It compares this prediction with the measurements. Doppler phase changes between resource times are compensated first. Departure and arrival directions agree for a direct monostatic route under the chosen conventions; different reflections on the outgoing and return routes can break this agreement.

Temporal coherence tests whether one Doppler shift explains phase changes across the CPI after antenna signals are combined for the estimated arrival direction. Amplitude and phase are fitted separately for each resource/polarization combination. Additional checks compare direction, speed, and power between the first and second CPI halves. A weaker, later return with the same direction and motion may follow another route to the same object. Counting it separately can duplicate a detection. These checks use measured consistency without knowing the true target locations or velocities.

Figure~\ref{fig:clutter} shows retained and rejected candidates. In this snapshot, fourteen energy candidates become six reports: five matched UAVs and one unmatched report. The wall example explains why a stable reflection can pass the checks. Several sites observing the same wall can even agree on a compatible virtual object. Coherence checks reject inconsistent paths, but passing them does not prove that the echo came directly from the UAV.

\subsection{False Outputs and Detection Coverage}
\begin{figure}[t]
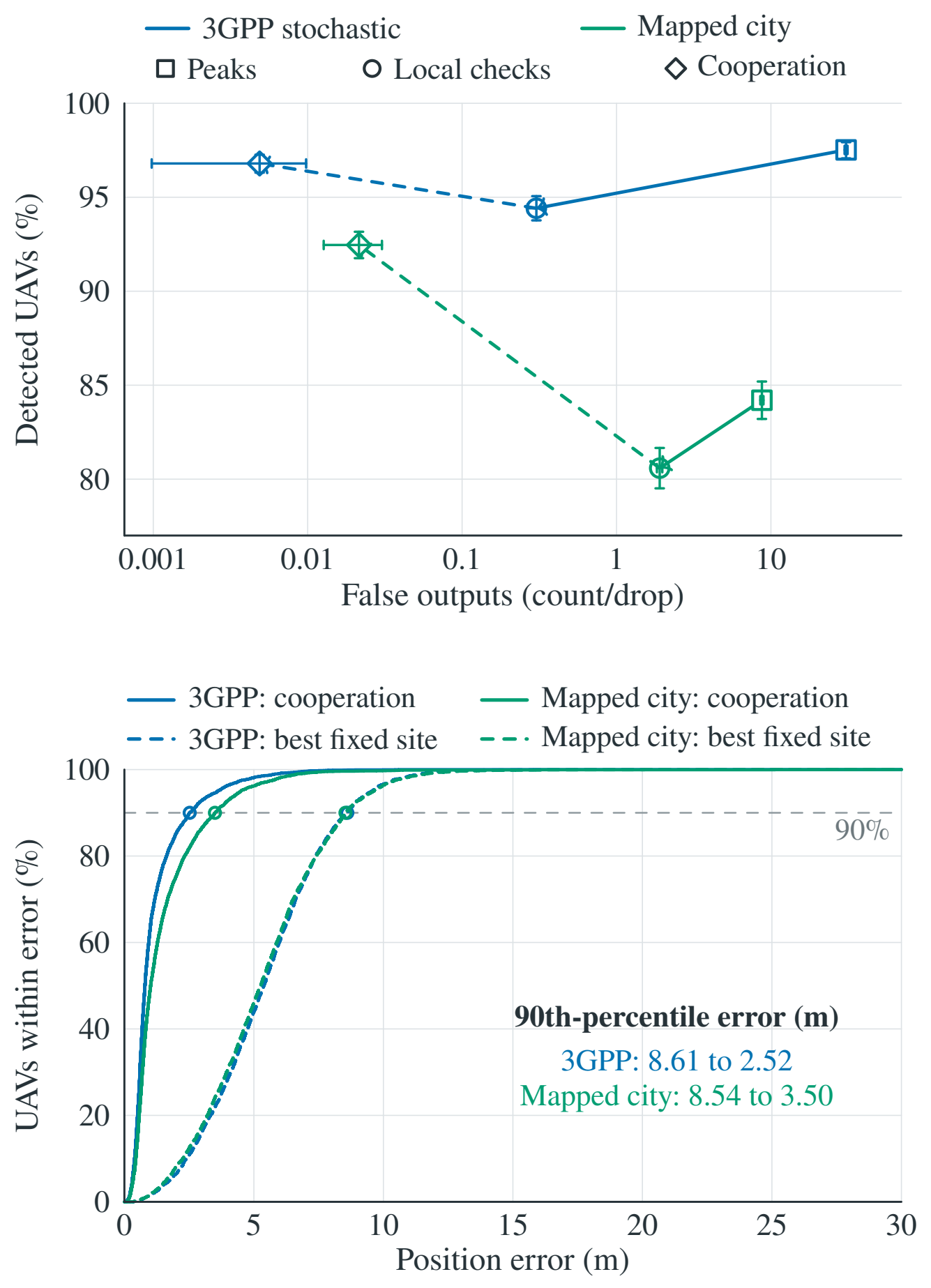

\centering
\definecolor{mrc0}{RGB}{221,226,229}
\definecolor{mrc1}{RGB}{38,50,56}
\definecolor{mrc2}{RGB}{0,114,178}
\definecolor{mrc3}{RGB}{0,158,115}

\makeatother

\caption{\textbf{Detection and positioning.} Upper: processing stages with 95\% bootstrap confidence intervals. Solid segments compare local checks on the same reference-site samples; dashed segments add three-site reception with the same total transmit resources and energy. Lower: cumulative position-error distributions compare the best fixed site and cooperation for UAVs detected by both methods.}
\label{fig:performance}
\end{figure}

The upper plot in Fig.~\ref{fig:performance} compares detection and false outputs across 1,024 occupied drops per environment, each containing five UAVs. Each output can match at most one UAV, and each UAV at most one output. Their positions must be within 30 m in three dimensions. Scoring first maximizes the number of matches, then minimizes their total position error. Unmatched outputs include ghosts, duplicates, and sufficiently displaced estimates. The horizontal axis counts unmatched outputs per occupied drop; the vertical axis gives the percentage of UAVs detected. Its energy-peak baseline already uses clutter subtraction, Hann weighting, noise-dependent thresholding, and peak separation; only the seven validation checks are disabled.

At the full-band reference site, the checks reduce mean unmatched outputs from 30.671 to 0.303 per stochastic drop and from 8.742 to 1.904 per mapped drop. Detection falls from 97.52\% to 94.41\% and from 84.20\% to 80.59\%, respectively. The checks remove many false outputs but also reduce the number of UAVs detected. Error bars indicate 95\% bootstrap confidence intervals obtained by resampling complete simulated scenes. They capture variation due to the finite evaluation population under fixed simulation assumptions. The three points compare processing stages rather than different detection thresholds.

Several checks can reject the same candidate. To account for this overlap, each check's added reduction is averaged over every possible order of enabling the seven checks. Spatial coherence, temporal coherence, and the transmit signature together account for about 83\% of removed stochastic false outputs and 84\% of mapped false outputs. Temporal coherence contributes most in the stochastic population; spatial coherence contributes most in the mapped population. Because checks overlap, these shares describe their average contributions; the ranking can change with the receiver and environment.

Overlapping reflections can reveal presence or motion; a known map can turn a reflected path into an additional location constraint. The evaluated checks favor the direct-return model. Rejecting every reflection would discard useful information, while treating every coherent reflection as a direct echo would create incorrect object locations. The design question is which paths can be interpreted reliably for the required observation.

\section{From Local Sensing to Network Knowledge}
\subsection{Complementary Views and Cooperative Decisions}
Separated sites provide different propagation paths~\cite{fishler_diversity} and different constraints on position and velocity~\cite{tang_uav}. Two sites viewing a target from almost the same direction may agree closely but reveal little about motion across that direction. A side view can reveal this missing motion component even when its echo is weaker. Choosing a sensing partner by power alone can therefore overlook useful geometry. The joint fit must check which position and velocity components these views reveal.

The network groups reports that could describe the same object and fits a shared position and velocity to each group. It checks agreement using range, direction, radial speed, site position and orientation, and measurement time. Reports with larger expected errors allow more disagreement; a systematic offset from reflection requires a physical explanation or a bias model. Strong reports pass all local checks; selected candidates that fail some checks are retained as weak reports. Agreement between strong reports supports a common object estimate. Weak reports can help when some sites lack a strong return. A report already assigned to one object cannot be reused for another.

At the cooperative point in the upper plot of Fig.~\ref{fig:performance}, three sites share the full-band single site's total reference-signal resources and energy at the transmitter output. They use 92/92/88 occupied blocks and 47.229 dBm per site, compared with 272 blocks and 52 dBm at one site. Cooperation also uses three receiving arrays. The comparison therefore changes both local processing and observation geometry; it does not isolate an algorithm on identical measurements. Cooperative detection reaches 96.80\% in the stochastic environment and 92.46\% in the mapped environment, with five and twenty-two unmatched outputs over the respective occupied populations.

For accuracy, the baseline is the full-band fixed site selected after evaluation for its highest overall detection rate. This differs from the reference site used to compare local checks. Only UAVs detected by both cooperation and this site are compared. Compared with this site, the 90th-percentile position error decreases from 8.61 to 2.52 m on 4,771 commonly detected stochastic targets and from 8.54 to 3.50 m on 4,511 commonly detected mapped targets. The lower plot in Fig.~\ref{fig:performance} shows cumulative distribution functions (CDFs): the fraction of these position errors no larger than the horizontal-axis value. No cooperative alarm was observed in 1,024 empty drops per environment, giving an approximately 0.29\% one-sided 95\% upper bound on the probability of an alarm in an empty drop.

\subsection{Reports That Retain Measurement Meaning}
The network needs more than a coordinate. Delay, direction, and Doppler need units and conventions; site identities, positions, and orientations establish the geometry. Timestamps give measurement times; ambiguity flags identify unresolved delay or speed alternatives. Local consistency scores describe how a candidate passed the checks. Candidate identifiers prevent reuse of the same report; keeping records for a known period allows later network requests.

Error covariance describes expected random errors and how they vary together; a bias model describes systematic offsets. Think of covariance as an uncertainty footprint. Range error stretches it along the viewing direction, while bearing error stretches it across that direction. A one-degree bearing error at 100 m produces about 1.75 m of transverse displacement; at 300 m it produces about 5.2 m. Tests of whether reports describe the same object should allow these unequal error spreads. Overlapping footprints support a match, but cannot rule out bias caused by a shared reflection. The evaluated fusion assigns weights using a model based on each report's signal-to-noise ratio and measurement-grid spacing. Measurements with larger modeled errors have less influence on the fit. The model assumes zero correlation between errors in different reports.

The same requirements apply when a base station illuminates an object and multiple user equipment devices receive bistatic returns. Reports must then include transmitter and receiver positions and orientations, together with their relative timing and frequency offsets. For moving receivers, their motion must also be known. An uncorrected 10 ns timing offset appears as approximately 3 m of extra bistatic path length; a frequency offset can be mistaken for motion. Reports should distinguish uncertainty in these offsets from uncertainty in the target measurement. More coordinate digits cannot correct an unknown clock offset. Local estimates can be fused without aligning carrier phase across receivers. Reports should distinguish the rate of path-length change measured by Doppler from the full velocity estimated using several views.

Compression can reduce sensing traffic~\cite{zhu_compression}. Instead of uploading every eligible weak report, receivers can retain them until requested. The network seeks reports compatible with strong candidates not yet assigned to an object. In the mapped study, selective requests reduce weak uploads from 1,558 to 213 but add 2,966 requests to individual receivers. Including strong uploads, modeled total traffic increases by 10.81\% when requests are as large as a measurement record and decreases by 4.03\% when they are one quarter that size. This accounting covers recovery of weak reports and excludes framing, acknowledgments, and requests for extra measurements to refine estimates. Reporting efficiency therefore depends on both uploads and requests.

\section{Design Lessons and Research Directions}
The value of network sensing lies in the decisions its observations enable. An air-corridor monitor needs timely awareness of an unexpected UAV; a road intersection needs complementary views around obstructions; a factory needs to monitor mobile robots and stationary obstacles. Each task needs different coverage, update time, and accuracy, and assigns different costs to missed objects and false reports. A useful 6G sensing service should choose signals, receiver processing, and sensing partners to meet each application's requirements.

The case study shows why these choices require physical interpretation. Removing unchanged channel responses reveals motion but can suppress an object the application still needs to observe. A coherent echo can still give the wrong location if a reflected path is treated as a direct route to the object. A view from another direction can reveal position or motion that a stronger echo from the original site cannot resolve. Statistical models reveal average tradeoffs; mapped propagation shows persistent ambiguities tied to surfaces. Together they guide designs for changing visibility and surroundings.

Several research directions follow. Tracking and adaptive illumination can direct observations toward objects whose uncertainty is growing. Map-aware processing can turn interpretable reflections into additional views. Measurement reporting can retain the timing, geometry, and uncertainty needed to combine observations across infrastructure and devices. Trials in operating networks should test these ideas with changing clutter, residual interference, and required response times. The opportunity for 6G is to make environmental awareness a coordinated network capability, with observation quality guiding when to sense, where to look, and when to seek another view.

\section*{Biography}
{\footnotesize\noindent\textbf{Xingqin Lin} is a Senior 3GPP Standards Engineer at NVIDIA. He received the IEEE ComSoc Fred W. Ellersick Prize and the IEEE VTS Early Career Award in 2021, and the IEEE ComSoc Best Young Professional Award in Industry in 2020. He is a Fellow of the Institution of Engineering and Technology. He was an IEEE ComSoc Distinguished Lecturer from 2024 to 2025 and an IEEE VTS Distinguished Lecturer from 2024 to 2026.\par}

\end{document}